\documentclass[letterpaper, 10 pt, conference]{ieeeconf}

\usepackage{amsfonts,amsmath,amssymb} 

\DeclareMathOperator{\diag}{diag}         
\def\rbb{\mathbb{R}}
\def\trp{^T}
\def\diag{{\rm diag}}

\def\half{\frac{1}{2}}
\usepackage{psfrag,color}
\usepackage{enumerate,cite,latexsym,graphicx}
\newtheorem{theorem}{Theorem}

\newtheorem{definition}{Definition}

\newtheorem{remark}{Remark}

\title{\LARGE \bf A Direct Coupling Coherent Quantum Observer}

\author{Ian R.~Petersen %
\thanks{This work was supported by the
Australian Research Council (ARC) and the Air Force Office of Scientific
Research (AFOSR). This material is based on research sponsored by the
Air Force Research Laboratory, under agreement number FA2386-12-1-4075.  The U.S. Government is authorized to reproduce and
distribute reprints for Governmental purposes notwithstanding any
copyright notation thereon.
The views and conclusions contained herein are those of the authors
and should not be interpreted as necessarily representing the official
policies or endorsements, either expressed or implied, of the Air
Force Research Laboratory or the U.S. Government. }%
\thanks{Ian R. Petersen is with the School of  Engineering and Information Technology, 
        University of New South Wales at the Australian Defence Force Academy, Canberra ACT 2600, Australia.
         {\tt\small i.r.petersen@gmail.com} } 
}%

\begin{document}

\maketitle
\thispagestyle{empty}
\pagestyle{empty}

\begin{abstract}
This paper considers the problem of constructing a direct coupling quantum observer for a closed linear quantum system. The proposed observer is shown to be able to estimate some but not all of the plant variables in a time averaged sense. A simple example and simulations are included to illustrate the properties of the observer.
\end{abstract}

\section{Introduction} \label{sec:intro}
A number of papers have recently considered the problem of constructing a coherent quantum observer for a quantum system; see \cite{MJ12a,VP9a,EMPUJ6a}. In the coherent quantum observer problem, a quantum plant is coupled to a quantum observer which is also a quantum system. The quantum observer is constructed to be a physically realizable quantum system  so that the system variables of the quantum observer converge in some suitable sense to the system variables of the quantum plant. 

In the papers  \cite{MJ12a,VP9a}, the quantum plant under consideration is a linear quantum system. In recent years, there has been considerable interest in the modeling and feedback control of linear quantum systems; e.g., see \cite{JNP1,NJP1,ShP5}.
Such linear quantum systems commonly arise in the area of quantum optics; e.g., see
\cite{GZ00,BR04}. For such linear quantum system models an important class of quantum control problems are referred to as coherent
quantum feedback control problems; e.g., see \cite{JNP1,NJP1,MaP3,MAB08,ZJ11,VP4,VP5a,HM12}. In these coherent quantum feedback control problems, both the plant and the controller are quantum systems and the controller is typically to be designed to optimize some performance index. The coherent quantum observer problem can be regarded as a special case of the coherent
quantum feedback control problem in which the objective of the observer is to estimate the system variables of the quantum plant. 

\begin{figure}[htbp]
\begin{center}
\includegraphics[width=8cm]{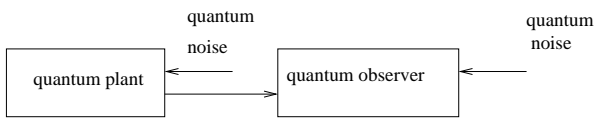}
\end{center}
\caption{Coherent Observer Structure with Field Coupling.}
\label{F1}
\end{figure}

In the previous papers on quantum observers such as  \cite{MJ12a,VP9a,EMPUJ6a}, the coupling between the plant and the observer is via a field coupling. This leads to an observer structure of the form shown in Figure \ref{F1}. This enables a one way connection between the quantum plant and the quantum observer. Also, since both the quantum plant and the quantum observer are  open quantum systems, they are both subject to quantum noise. 

\begin{figure}[htbp]
\begin{center}
\includegraphics[width=8cm]{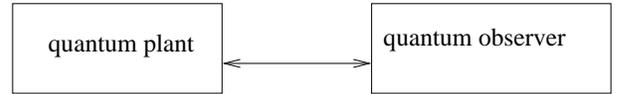}
\end{center}
\caption{Coherent Observer Structure with Direct Coupling.}
\label{F2}
\end{figure}

However in the paper \cite{ZJ11}, a coherent quantum control problem is considered in which both field coupling and direct coupling is considered between the quantum plant and the quantum controller. In this paper, we explore the construction of a coherent quantum observer in which there is only direct coupling between quantum plant and the quantum observer. Furthermore, both the quantum plant and the quantum observer are assumed to be closed quantum systems which means that they are not subject to quantum noise and are purely deterministic systems. This leads to an observer structure of the form shown in Figure \ref{F2}. It is shown that for the case being considered, a quantum observer can be constructed to estimate some but not all of the system variables of the quantum plant. Also, the observer variables converge to the plant variables in a time averaged sense rather than a quantum expectation sense such as considered in the papers \cite{MJ12a,VP9a}.

\section{Quantum Linear Systems}
In this section, we describe the  class of closed linear quantum systems under consideration; see also \cite{JNP1,GJ09,ZJ11}. We consider linear non-commutative systems of the form
\begin{eqnarray}
\dot x(t) &=& Ax(t); \quad x(0)=x_0
 \label{linear-c}
\end{eqnarray}
where $A$ is a real matrix in $\rbb^{n
\times n}$, and $ x(t) = [\begin{array}{ccc} x_1(t) & \ldots &
x_n(t)
\end{array}]\trp$ is a vector of self-adjoint possibly
non-commutative system variables; e.g., see \cite{JNP1}. Here $n$ is assumed to be an even number and $\frac{n}{2}$ is the number of modes in the quantum system. 

The initial system variables $x(0)=x_0$ 
are assumed to satisfy the {\em commutation relations}
\begin{equation}
[x_j(0), x_k(0) ] = 2 i \Theta_{jk}, \ \ j,k = 1, \ldots, n,
\label{x-ccr}
\end{equation}
where $\Theta$ is a real antisymmetric matrix with components
$\Theta_{jk}$.  Here, the commutator is defined
by $[A,B]=AB-BA$. In the case of a
single degree of freedom quantum particle, $x=(x_1, x_2)^T$ where
$x_1=q$ is the position operator, and $x_2=p$ is the momentum
operator.  The
commutation relations are  $[q,p]=2 i$.
Here, the matrix $\Theta$ is assumed to be  of the  form
 $\Theta=\diag(J,J,\ldots,J)$ where $J$ denotes the real skew-symmetric $2\times 2$ matrix
$$
J= \left[ \begin{array}{cc} 0 & 1 \\ -1 & 0
\end{array} \right].$$

A linear quantum  system
(\ref{linear-c}) is said to be \emph{physically realizable} if it ensures 
the preservation of the canonical commutation relations (CCRs):
$$x(t)x(t)\trp-(x(t)x(t)\trp)\trp=2i\Theta \ \mbox{for all } t\geq 0.$$
This holds when the system (\ref{linear-c}) corresponds to a collection of \emph{closed quantum
harmonic oscillators}; see \cite{JNP1}.  Such
quantum harmonic oscillators are  described by a quadratic Hamiltonian
$\mathcal{H}=\half x(0)\trp R x(0)$, where $R$ is a real symmetric matrix. 

\begin{theorem}[\cite{JNP1}]  \label{thm_phys_real}
The system (\ref{linear-c}) is physically realizable if and only
if:
\begin{eqnarray}
A\Theta + \Theta A \trp&=&0. 
\label{eq_realize_cond_A}
\end{eqnarray}
In this case, the corresponding  Hamiltonian matrix $R$ is  given by
$R=\frac{1}{4}(-\Theta A+A\trp \Theta)$. In addition, for a given Hamiltonian matrix $R$, the corresponding matrix $A$ in 
(\ref{linear-c}) is given by 
\begin{equation}
A=2\Theta R \label{eq_coef_cond_A}.
\end{equation}
\end{theorem}

\begin{remark}
\label{R1}
Note that the system (\ref{linear-c}) cannot be asymptotically stable if it is physically realizable. To see this, first suppose $R \neq 0$. Then, observe that the Hamiltonian is preserved in time. Indeed,
$ \mathcal{\dot H} = \frac{1}{2}\dot{x}^TRx+\frac{1}{2}x^TR\dot{x} = -x^TR\Theta R x + x^TR\Theta R x = 0$ since $R$ is symmetric and $\Theta$ is skew-symmetric. However, if the system were asymptotically stable, then $x(t) \rightarrow 0$ as $t \rightarrow \infty$ which would contradict this fact. Also, if $R=0$, then $A = 0$ which is again not asymptotically stable. A similar conclusion can also be drawn from the fact that the CCRs are preserved in time. 

Since it is not possible for a physically realizable quantum system of the form (\ref{linear-c}) to be asymptotically stable, we will need a new notion of convergence for our direct coupled quantum observer. 
\end{remark}

\section{Direct Coupling Coherent Quantum Observers}
We first consider general {\em closed linear quantum plants} described by
non-commutative models of the following form:
\begin{eqnarray}
\dot x_p(t) &=& A_px_p(t); \quad x_p(0)=x_{0p}; \nonumber \\
z_p(t) &=& C_px_p(t)
 \label{plant}
\end{eqnarray}
where $z_p$ denotes the vector of system variables to be estimated by the observer and $ A_p \in \rbb^{n_p
\times n_p}$, $C_p\in \rbb^{m_p \times n_p}$. 
It is assumed that this quantum plant is physically realizable and corresponds to a plant Hamiltonian
$\mathcal{H}_p=\half x_p(0)\trp R_p x_p(0)$ where the symmetric matrix $R_p$ is given by
$R_p = \frac{1}{4}(-\Theta A_p+A_p\trp \Theta)$. 

Also, we consider a {\em direct coupled linear quantum observer} defined by a symmetric matrix $R_o \in \rbb^{n_o
\times n_o}$, and matrices $R_c \in \rbb^{n_p \times n_o}$, $C_o\in \rbb^{m_p \times n_o}$. These matrices define an observer Hamiltonian
\begin{equation}
\label{observer_hamiltonian}
\mathcal{H}_o=\half x_o(0)\trp R_o x_o(0),
\end{equation}
and a coupling Hamiltonian
\begin{equation}
\label{coupling_hamiltonian}
\mathcal{H}_c=\half x_p(0)\trp R_c x_o(0)+\half x_o(0)\trp R_c\trp x_p(0).
\end{equation}
The matrix $C_o$ also defines the vector of estimated variables for the observer as $z_o(t)  = C_ox_o(t)$. 

The augmented quantum linear system consisting of the quantum plant and the direct coupled  quantum observer is then a quantum system of the form (\ref{linear-c}) described by the total Hamiltonian
\begin{eqnarray}
\mathcal{H}_a &=& \mathcal{H}_p+\mathcal{H}_c+\mathcal{H}_o\nonumber \\
 &=& \half x_a(0)\trp R_a x_a(0)
\label{total_hamiltonian}
\end{eqnarray}
where
$x_a = \left[\begin{array}{l}x_p\\x_o\end{array}\right]$ and 
$R_a = \left[\begin{array}{ll}R_p & R_c\\R_c^T & R_o\end{array}\right]$. Then, using (\ref{eq_coef_cond_A}), it follows that the augmented quantum linear system is described by the equations
\begin{eqnarray}
\left[\begin{array}{l}\dot x_p(t)\\\dot x_o(t)\end{array}\right] &=& 
A_a\left[\begin{array}{l} x_p(t)\\ x_o(t)\end{array}\right];~ x_p(0)=x_{0p};~ x_o(0)=x_{0o};\nonumber \\
z_p(t) &=& C_px_p(t);\nonumber \\
z_o(t) &=& C_ox_o(t)
\label{augmented_system}
\end{eqnarray}
where $A_a = 2\Theta R_a$. 

We now formally define the notion of a direct coupled linear quantum observer.

\begin{definition}
\label{D1}
The matrices $R_o \in \rbb^{n_o
\times n_o}$,  $R_c \in \rbb^{n_p \times n_o}$, $C_o\in \rbb^{m_p \times n_o}$ define a {\em direct coupled linear quantum observer} for the quantum linear plant (\ref{plant}) if the corresponding augmented linear quantum system (\ref{augmented_system}) is such that
\begin{equation}
\label{average_convergence}
\lim_{T \rightarrow \infty} \frac{1}{T}\int_{0}^{T}(z_p(t) - z_o(t))dt = 0.
\end{equation}
\end{definition}

\begin{remark}
\label{R0}
Note that although the direct coupling coherent quantum observer defined above does not use field coupling to connect the quantum observer to the quantum plant, quantum optics may be used in order to physically realize the augmented plant-observer system (\ref{augmented_system}). Indeed, using the methods proposed in the papers \cite{NUR10,NJD09,NUR10A,PET10Ca,PET08A}, the augmented system could be physically realized using quantum optics without the use of direct couplings between modes but rather using internal field couplings; see also \cite{GJ09}.
\end{remark} 

\section{Constructing a Direct Coupling Coherent Quantum Observer}
We now describe the construction of a direct coupled linear quantum observer.  In this section, we assume that  $A_p =0$ in (\ref{plant}). This corresponds to $R_p = 0$ in the plant Hamiltonian. It follows from (\ref{plant}) that the plant system variables $x_p(t)$ will remain fixed if the plant is not coupled to the observer. However, when the plant is coupled to the quantum observer this will no longer be the case. We will show that if the quantum observer is suitably designed, the plant quantity to be estimated  $z_p(t)$ will remain fixed and the condition (\ref{average_convergence}) will be satisfied. 

We also assume that $m_p = \frac{n_p}{2}$ and the matrix $C_p$ is of the form $C_p = \beta^T$ where
\begin{equation}
\label{beta}
\beta = \left[\begin{array}{llll}\beta_1 & 0 & & \\
0 & \beta_2  & & 0\\
&  & \ddots & \\
& 0 & & \beta_{\frac{n_p}{2}}
\end{array}\right] \in \rbb^{n_p \times \frac{n_p}{2}}
\end{equation}
and $\beta_i \in \rbb^{2\times 1}$ for $i=1,2,\ldots,\frac{n_p}{2}$. This assumption means that the plant variables to be estimated include only one quadrature for each mode of the plant. 

We now suppose that  the matrices $R_o$,  $R_c$, $C_o$ are such that $R_c = \beta \alpha^T$, $\alpha \in \rbb^{n_o \times \frac{n_p}{2}}$ and the matrix $R_o$ is positive definite. Also, we write 
$\Theta =  \left[\begin{array}{ll}\Theta_1 & 0\\0 & \Theta_2\end{array}\right]$ where $\Theta_1 \in \rbb^{n_p\times n_p}$ and 
$\Theta_2 \in \rbb^{n_o\times n_o}$. Then, 
$R_a = \left[\begin{array}{ll}0 & \beta \alpha^T\\\alpha \beta^T & R_o\end{array}\right]$ and
$A_a = 2\Theta R_a = \left[\begin{array}{ll}0 & 2\Theta_1 \beta \alpha^T\\2 \Theta_2\alpha \beta^T & 2 \Theta_2R_o\end{array}\right]$. Hence, the augmented system equations (\ref{augmented_system}) describing the combined plant-observer system become
\begin{eqnarray}
\dot x_p(t)&=& 2\Theta_1 \beta \alpha^Tx_o(t);\nonumber \\
\dot x_o(t)&=&2 \Theta_2\alpha \beta^Tx_p(t)+2 \Theta_2R_ox_o(t);\nonumber \\
z_p(t) &=& C_px_p(t);\nonumber \\
z_o(t) &=& C_ox_o(t). 
\label{augmented_system1}
\end{eqnarray}
We now use Laplace Transforms to solve these equations. It follows that
\begin{eqnarray}
\label{laplace_aug_sys}
 sX_p(s) &=& 2\Theta_1 \beta \alpha^TX_o(s)+x_p(0);\nonumber \\
sX_o(s)&=&2 \Theta_2 \alpha \beta^TX_p(s)+2 \Theta_2R_oX_o(s) +x_o(0) \nonumber \\
\end{eqnarray}
and hence,
\begin{eqnarray*}
sX_o(s)&=&\frac{4}{s} \Theta_2\alpha \beta^T\Theta_1 \beta \alpha^TX_o(s)+\frac{2}{s} \Theta_2\alpha \beta^Tx_p(0)\\
&&+2 \Theta_2R_oX_o(s) +x_o(0).
\end{eqnarray*}
However, 
\[
\beta^T\Theta_1 \beta = 
\left[\begin{array}{llll}\beta_1^TJ\beta_1 & 0 & & \\
0 & \beta_2^TJ\beta_2  & & 0\\
&  & \ddots & \\
& 0 & & \beta_{n_p}^TJ\beta_{n_p}
\end{array}\right] = 0
\]
 since $J$ is skew-symmetric. Therefore,
\begin{equation}
\label{Xos}
X_o(s) = \left(sI-2 \Theta_2R_o\right)^{-1}\left(\frac{2}{s} \Theta_2\alpha \beta^Tx_p(0)+x_o(0)\right).
\end{equation}
Taking the inverse Laplace Transform of this equation, we obtain
\begin{eqnarray}
\label{xot}
x_o(t) &=& e^{2\Theta_2R_ot}x_o(0)+2\int_o^te^{2\Theta_2R_o(t-\tau)}d\tau \Theta_2\alpha \beta^Tx_p(0)\nonumber\\
&=& e^{2\Theta_2R_ot}x_o(0)\nonumber \\
&&-e^{2\Theta_2R_ot}\left(e^{-2\Theta_2R_ot} - I\right)R_o^{-1}\Theta_2^{-1}\Theta_2\alpha \beta^Tx_p(0)\nonumber\\
&=&e^{2\Theta_2R_ot}\left(x_o(0) + R_o^{-1}\alpha \beta^Tx_p(0)\right)\nonumber\\
&&-R_o^{-1}\alpha \beta^Tx_p(0).
\end{eqnarray}

Also, if we substitute (\ref{Xos}) into (\ref{laplace_aug_sys}), we obtain
\begin{eqnarray*}
\label{Xps}
X_p(s) 
&=&\frac{4}{s^2}\Theta_1 \beta \alpha^T\left(sI-2 \Theta_2R_o\right)^{-1}\Theta_2\alpha \beta^Tx_p(0)\\
&&+\frac{2}{s} \Theta_1 \beta \alpha^T\left(sI-2 \Theta_2 R_o\right)^{-1}x_o(0)\\
&&+\frac{1}{s}x_p(0).
\end{eqnarray*}
Taking the inverse Laplace Transform of this equation, we obtain
\begin{eqnarray}
\label{xpt}
x_p(t) &=&4\Theta_1 \beta \alpha^T\int_o^te^{2\Theta_2R_o(t-\tau)}\tau d\tau \Theta_2\alpha \beta^Tx_p(0)\nonumber \\
&&+2\Theta_1 \beta \alpha^T\int_o^te^{2\Theta_2R_o(t-\tau)}d\tau x_o(0)\nonumber \\
&&+x_p(0)\nonumber \\
&=& -2t\Theta_1 \beta \alpha^TR_o^{-1}\alpha \beta^Tx_p(0)\nonumber \\
&&+\Theta_1 \beta \alpha^T R_o^{-2}\Theta_2\alpha \beta^Tx_p(0)\nonumber \\
&&-\Theta_1 \beta \alpha^T e^{2\Theta_2 R_ot}R_o^{-2}\Theta_2 \alpha \beta^Tx_p(0)\nonumber \\
&&+\Theta_1 \beta \alpha^TR_o^{-1}\Theta_2x_o(0)\nonumber \\
&&-\Theta_1 \beta \alpha^Te^{2\Theta_2R_ot}R_o^{-1}\Theta_2x_o(0)\nonumber \\
&&+x_p(0).
\end{eqnarray}
We now choose the parameters of the quantum observer so that  $C_oR_o^{-1}\alpha = -I$. It follows from (\ref{xot}) and (\ref{xpt}) that the quantities $z_p(t) = C_px_p(t)$ and $z_o(t) = C_ox_o(t)$ are given by
\begin{equation}
\label{zot}
z_o(t) = C_oe^{2\Theta_2R_ot}\left(x_o(0) + R_o^{-1}\alpha \beta^Tx_p(0)\right)+z_p(0)
\end{equation}
and
\begin{equation}
\label{zop}
z_p(t) = z_p(0)
\end{equation}
where we have used the fact that $C_p\Theta_1 \beta = \beta^T\Theta_1 \beta =0$. That is, the quantity $z_p(t)$ remains constant and is not affected by the coupling to the coherent quantum observer. 

Note that the equation (\ref{zop}) can be derived directly since 
\begin{eqnarray*}
\left[\begin{array}{ll}C_p & 0\end{array}\right]A_a 
&=& \left[\begin{array}{ll}\beta^T & 0\end{array}\right]
\left[\begin{array}{ll}0 & 2\Theta_1 \beta \alpha^T\\2 \Theta_2\alpha \beta^T & 2 \Theta_2 R_o\end{array}\right] \nonumber \\
&=& \left[\begin{array}{ll}0 & 2\beta^T\Theta_1 \beta \alpha^T\end{array}\right] \nonumber \\
&=& 0
\end{eqnarray*}
since $\beta^T\Theta_1 \beta =0$. Hence,
\[
z_p(t) = \left[\begin{array}{ll}C_p & 0\end{array}\right]e^{A_at}x_a(0) = \left[\begin{array}{ll}C_p & 0\end{array}\right]x_a(0) = z_p(0)
\]
for all $t \geq 0$. 

Note that the matrix $A_a$ will have all purely imaginary eigenvalues. To see this, we first observe that the matrix $2i\Theta_2 R_o$ has 
purely real eigenvalues since  $2i\Theta_2$ is a Hermitian matrix  and $R_o$ is assumed to be a positive definite matrix. Indeed, 
$2i\Theta_2 R_o = 2R_o^{-\half}R_o^{\half}\Theta_2 R_o^{\half}R_o^{\half}$ and thus $2i\Theta_2 R_o$ is similar to the Hermitian matrix 
$2iR_o^{\half}\Theta_2 R_o^{\half}$ which has purely real eigenvalues. Hence, $2\Theta_2 R_o$ must have purely imaginary eigenvalues. 

Now suppose the vector $\left[\begin{array}{l}x_p\\x_o\end{array}\right]$ is an eigenvector of $A_a$ with corresponding eigenvector $\lambda$. 
Hence,
\[
\left[\begin{array}{ll}0 & 2\Theta_1 \beta \alpha^T\\2 \Theta_2\alpha \beta^T & 2 \Theta_2R_o\end{array}\right]\left[\begin{array}{l}x_p\\x_o\end{array}\right] = \lambda \left[\begin{array}{l}x_p\\x_o\end{array}\right]
\]
and hence
\begin{equation}
\label{eq1}
2\Theta_1 \beta \alpha^Tx_o=\lambda x_p
\end{equation}
and
\begin{equation}
\label{eq2}
2 \Theta_2\alpha \beta^Tx_p+ 2 \Theta_2 R_ox_o =\lambda x_o.
\end{equation}
We now premultiply (\ref{eq1}) by $\beta^T$ and use the fact that $\beta^T\Theta_1 \beta =0$ to obtain
\[
\lambda \beta^T x_p = 0.
\]
Hence, either $\lambda = 0$ which means that the eigenvalue is purely imaginary or $\beta^T x_p = 0$. If $\lambda \neq 0$ the condition $\beta^T x_p = 0$ is substituted into (\ref{eq2}) to obtain
\[
2 \Theta_2 R_ox_o =\lambda x_o.
\]
Furthermore, if $x_o=0$, it follows from (\ref{eq1}) that $\lambda x_p = 0$ and hence, $ x_p = 0$ since $\lambda \neq 0$. However, this contradicts the fact that $\left[\begin{array}{l}x_p\\x_o\end{array}\right]$ is an eigenvector of $A_a$. Thus, we must have $x_o\neq 0$. Thus, we can now conclude that $\lambda$ is an eigenvalue of $2 \Theta_2 R_o$ which we have already established has only purely imaginary eigenvalues. Thus, $\lambda$ must be purely imaginary in this case as well.

We now verify that the condition (\ref{average_convergence}) is satisfied for this quantum observer. We recall from Remark \ref{R1} that the quantity $\half x\trp R_o x$
remains constant in time for the linear system:
\[
\dot x = 2\Theta_2R_o x;\quad x(0) = x_0.
\]
That is 
\begin{equation}
\label{Roconst}
\half x(t) \trp R_o x(t) = \half x_0 \trp R_o x_0 \quad \forall t \geq 0.
\end{equation}
However, $x(t) = e^{2\Theta_2R_ot}x_0$ and $R_o > 0$. Therefore, it follows from (\ref{Roconst}) that
\[
\sqrt{\lambda_{min}(R_o)}\|e^{2\Theta_2R_ot}x_0\| \leq \sqrt{\lambda_{max}(R_o)}\|x_0\|
\]
for all $x_0$ and $t \geq 0$. Hence, 
\begin{equation}
\label{exp_bound}
\|e^{2\Theta_2R_ot}\| \leq \sqrt{\frac{\lambda_{max}(R_o)}{\lambda_{min}(R_o)}}
\end{equation}
for all $t \geq 0$.

Now since $\Theta_2$ and $R_o$ are non-singular,
\[
\int_0^Te^{2\Theta_2R_ot}dt = \half e^{2\Theta_2R_oT}R_o^{-1}\Theta_2^{-1} - \half R_o^{-1}\Theta_2^{-1}
\]
and therefore, it follows from (\ref{exp_bound}) that
\begin{eqnarray*}
\lefteqn{\frac{1}{T} \|\int_0^Te^{2\Theta_2R_ot}dt\|}\nonumber \\
 &=& \frac{1}{T} \|\frac{1}{2}e^{2\Theta_2R_oT}R_o^{-1}\Theta_2^{-1} - \frac{1}{2}R_o^{-1}\Theta_2^{-1}\|\nonumber \\
&\leq& \frac{1}{2T}\|e^{2\Theta_2R_oT}\|\|R_o^{-1}\Theta_2^{-1}\| \nonumber \\
&&+ \frac{1}{2T}\|R_o^{-1}\Theta_2^{-1}\|\nonumber \\
&\leq&\frac{1}{2T}\sqrt{\frac{\lambda_{max}(R_o)}{\lambda_{min}(R_o)}}\|R_o^{-1}\Theta_2^{-1}\|\nonumber \\
&&+\frac{1}{2T}\|R_o^{-1}\Theta_2^{-1}\|\nonumber \\
&\rightarrow & 0 
\end{eqnarray*}
as $T \rightarrow \infty$. Hence, (\ref{zot}) implies
\[
\lim_{T \rightarrow \infty} \frac{1}{T}\int_{0}^{T} z_o(t)dt = z_p(0).
\]
Also, (\ref{zop}) implies 
\[
\lim_{T \rightarrow \infty} \frac{1}{T}\int_{0}^{T} z_p(t)dt = z_p(0).
\]
Therefore, condition (\ref{average_convergence}) is satisfied. Thus, we have established the following theorem.

\begin{theorem}
\label{T2}
Consider a quantum plant of the form (\ref{plant}) where  $A_p = 0$, $C_p = \beta^T$ and $\beta$ is as defined in (\ref{beta}). Then  the matrices $R_o>0$, $R_c$, $C_o$ will define direct coupled quantum observer for this quantum plant if $R_c$ is of the form $R_c = C_p^T\alpha^T$ where $\alpha \in \rbb^{n_o \times \frac{n_p}{2}}$ and $C_o^TR_o^{-1}\alpha = -I$.
\end{theorem}

\begin{remark}
\label{R2}
 We consider the above result for the single mode case with $n_p =2$, $m_p=1$, in which $C_p = [1~0]$. This means that the variable to be estimated by the quantum observer is the position operator of the quantum plant; i.e., $z_p(t) = q_p(t)$ where 
$x_p(t) = \left[\begin{array}{l}q_p(t)\\p_p(t)\end{array}\right]$. By choosing $n_o = 2$, $R_o = I$, $C_o = [1~0]$, $\beta = \left[\begin{array}{l}1\\0\end{array}\right]$ and $\alpha = \left[\begin{array}{l}-1\\0\end{array}\right]$, the conditions of  Theorem \ref{T2} will be satisfied and the observer output variable will be the position operator of the quantum observer $q_o(t)$; i.e., $z_o(t) = q_o(t)$ where 
$x_o(t) = \left[\begin{array}{l}q_o(t)\\p_o(t)\end{array}\right]$.  Before the quantum observer is connected to the quantum plant, the quantities $q_p(t)$ and $p_p(t)$ will remain constant since we have assumed that $A_p = 0$. Now suppose that the quantum observer is connected to the quantum plant at time $t = 0$. According to (\ref{zop}), the plant position operator $q_p(t)$ will remain constant at its initial value $q_p(t)=q_p(0)$ but the plant momentum operator $p_p(t)$ will evolve in an time varying and oscillatory way as defined by (\ref{xpt}). In addition, the observer position operator $q_o(t)$ will evolve in an oscillatory way as defined by (\ref{zot}) but its time average will converge to $q_p(0)$ according to (\ref{average_convergence}). 

Now suppose that after a sufficiently long time $T$ such that the time average of $q_o(t)$ has essentially converged to $q_p(0)$, the observer is disconnected from the quantum plant. Then, the plant position operator $q_p(t)$ will remain constant at $q_p(t)=q_p(0)$ and the plant momentum operator $p_p(t)$ will remain constant at a value $p_p(T)$ which is determined by the formula (\ref{xpt}) in terms of $x_p(0)$, $x_o(0)$ and the time $T$. This will be an essentially random value. If at a later time an observer with the same parameters as above is connected to the quantum plant, then time average of its output $z_o(t) = q_o(t)$ will again converge to $q_p(0)$ and $q_p(t)$ will remain constant at $q_p(t)=q_p(0)$. However, suppose that instead an observer with different parameters $R_o = I$, $C_o = [0~1]$ and $\alpha = \left[\begin{array}{l}0\\-1\end{array}\right]$ 
is used.
This observer is designed so that the time average of the observer output $z_o(t) = p_o(t)$ converges to the momentum operator of the quantum plant $p_p(t)$. This quantity is the essentially random value $p_p(T)$ mentioned above. In addition, the previously constant value of $q_p(t)=q_p(0)$ will now be destroyed and will evolve to another essential random value. This behavior of the quantum observer is similar to the behavior of quantum measurements; e.g., see \cite{WM10}. This is not surprising since the behavior of the direct coupled quantum observers considered in this paper and the behavior of quantum measurements are both determined by the quantum commutation relations which are fundamental to the theory of quantum mechanics.
\end{remark}

\section{Numerical Simulations of a Quantum Observer for a One Mode Plant}
\begin{figure}[htbp]
\begin{center}
\includegraphics[width=7cm]{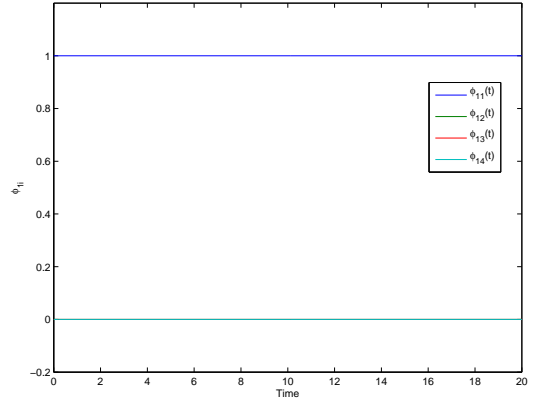}
\end{center}
\caption{Coefficient functions defining $q_p(t)$.}
\label{F3}
\end{figure}

We now present some numerical simulations to illustrate the direct coupled quantum observer described in the previous section. We consider the quantum observer considered in Remark \ref{R2} above where $n_p = 2$, $m_p =1$, $n_o = 2$, $A_p = 0$, $C_p = [1~0]$, $R_o = I$, $C_o = [1~0]$, 
$\beta = \left[\begin{array}{l}1\\0\end{array}\right]$ and  $\alpha = \left[\begin{array}{l}-1\\0\end{array}\right]$. As described in Remark \ref{R2}, the variable to be estimated by the quantum observer is the position operator of the quantum plant; i.e., $z_p(t) = q_p(t)$ where $x_p(t) = \left[\begin{array}{l}q_p(t)\\p_p(t)\end{array}\right]$. Also, the observer output variable will be the position operator of the quantum observer $q_o(t)$; i.e., $z_o(t) = q_o(t)$ where 
$x_o(t) = \left[\begin{array}{l}q_o(t)\\p_o(t)\end{array}\right]$. Then the augmented plant-observer system is described by the equations
\[
\left[\begin{array}{l}\dot q_p(t)\\ \dot p_p(t)\\ \dot q_o(t)\\\dot p_o(t)\end{array}\right]
= A_a \left[\begin{array}{l} q_p(t)\\  p_p(t)\\  q_o(t)\\ p_o(t)\end{array}\right]
\]
where
\[
A_a = \left[\begin{array}{ll}0 & 2J \beta \alpha^T\\2 J\alpha \beta^T & 2 JR_o\end{array}\right]
= \left[\begin{array}{llll}      0  &   0  &  0  &   0 \\
     0 &    0  &   2  &   0 \\
    0 &    0  &   0  &   2 \\
     2 &    0  &   -2  &  0
\end{array}\right].
\]
Then, we can write
\[
\left[\begin{array}{l} q_p(t)\\  p_p(t)\\  q_o(t)\\ p_o(t)\end{array}\right] 
= \Phi(t) \left[\begin{array}{l} q_p(0)\\  p_p(0)\\  q_o(0)\\ p_o(0)\end{array}\right]
\]
where 
\[
\Phi(t) = \left[\begin{array}{llll}      \phi_{11}(t)  &   \phi_{12}(t)  &  \phi_{13}(t)  &  \phi_{14}(t) \\
     \phi_{21}(t)  &   \phi_{22}(t)  &  \phi_{23}(t)  &  \phi_{24}(t) \\
\phi_{31}(t)  &   \phi_{32}(t)  &  \phi_{33}(t)  &  \phi_{34}(t) \\
\phi_{41}(t)  &   \phi_{42}(t)  &  \phi_{43}(t)  &  \phi_{44}(t)
\end{array}\right]
= e^{A_a t}.
\]
Thus, the plant variable to be estimated $q_p(t)$ is given by
\[
q_p(t) = \phi_{11}(t)q_p(0)+\phi_{12}(t)p_p(0)+\phi_{13}(t)q_o(0)+\phi_{14}(t)p_o(0)
\]
and we plot the functions $\phi_{11}$, $\phi_{12}(t)$, $\phi_{13}(t)$, $\phi_{14}(t)$ in Figure \ref{F3}. 
From this figure, we can see that $\phi_{11}(t)\equiv 1$, $\phi_{12}(t)\equiv 0$, $\phi_{13}(t)\equiv 0$, $\phi_{14}(t)\equiv 0$, and $q_p(t)$ will remain constant at $q_p(0)$ for all $t \geq 0$. 

\begin{figure}[htbp]
\begin{center}
\includegraphics[width=7cm]{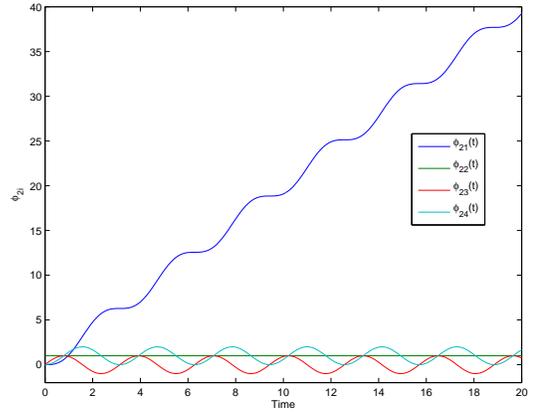}
\end{center}
\caption{Coefficient functions defining $p_p(t)$.}
\label{F4}
\end{figure}

Also, the other plant variable  $p_p(t)$ is given by
\[
p_p(t) = \phi_{21}(t)q_p(0)+\phi_{22}(t)p_p(0)+\phi_{23}(t)q_o(0)+\phi_{24}(t)p_o(0)
\]
and we plot the functions $\phi_{21}$, $\phi_{22}(t)$, $\phi_{23}(t)$, $\phi_{24}(t)$ in Figure \ref{F4}. 
From this figure, we can see that $p_p(t)$  evolves in a time-varying and oscillatory way when the quantum plant is connected to the quantum observer. 

\begin{figure}[htbp]
\begin{center}
\includegraphics[width=7cm]{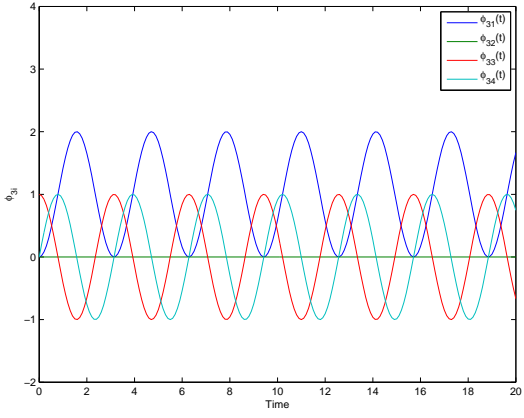}
\end{center}
\caption{Coefficient functions defining $q_o(t)$.}
\label{F5}
\end{figure}

\begin{figure}[htbp]
\begin{center}
\includegraphics[width=7cm]{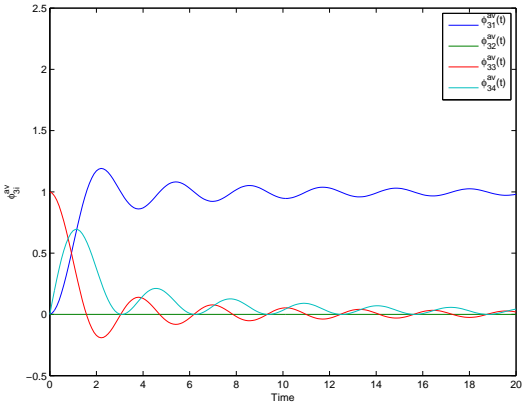}
\end{center}
\caption{Coefficient functions defining the time average of  $q_o(t)$.}
\label{F6}
\end{figure}

We now consider the output variable of the quantum observer $q_o(t)$ which is given by
\[
q_o(t) = \phi_{31}(t)q_p(0)+\phi_{32}(t)p_p(0)+\phi_{33}(t)q_o(0)+\phi_{34}(t)p_o(0)
\]
and we plot the functions $\phi_{31}$, $\phi_{32}(t)$, $\phi_{33}(t)$, $\phi_{34}(t)$ in Figure \ref{F5}.
To illustrate the time average convergence property of the quantum observer (\ref{average_convergence}), we now plot the quantities
\begin{eqnarray*}
\phi_{31}^{ave}(T) &=& \frac{1}{T}\int_0^T\phi_{31}(t)dt\nonumber \\ 
\phi_{32}^{ave}(T) &=& \frac{1}{T}\int_0^T\phi_{32}(t)dt\nonumber \\ 
\phi_{33}^{ave}(T) &=& \frac{1}{T}\int_0^T\phi_{33}(t)dt\nonumber \\ 
\phi_{34}^{ave}(T) &=& \frac{1}{T}\int_0^T\phi_{34}(t)dt\nonumber \\ 
\end{eqnarray*}
in Figure \ref{F6}. 
From this figure, we can see that the time average of $q_o(t)$  converges to $q_p(0)$ as $t \rightarrow \infty$. Note that the effect of time averaging can be regarded as a low pass filtering effect which removes the sinusoidal oscillations but retains the DC component which represents the estimate of the plant variable. 

\begin{figure}[htbp]
\begin{center}
\includegraphics[width=7cm]{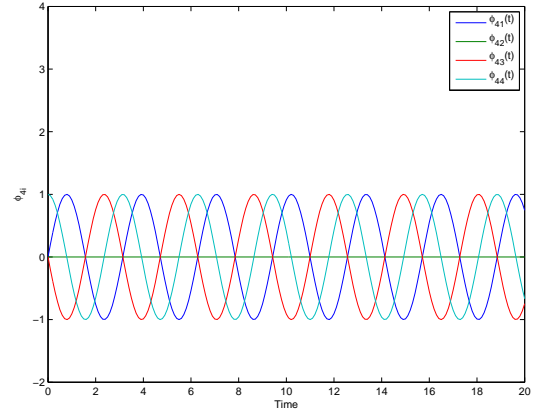}
\end{center}
\caption{Coefficient functions defining $p_o(t)$.}
\label{F6a}
\end{figure}

We now consider the other variable of the quantum observer $p_o(t)$ which is given by
\[
p_o(t) = \phi_{41}(t)q_p(0)+\phi_{42}(t)p_p(0)+\phi_{43}(t)q_o(0)+\phi_{44}(t)p_o(0)
\]
and we plot the functions $\phi_{41}$, $\phi_{42}(t)$, $\phi_{43}(t)$, $\phi_{44}(t)$ in Figure \ref{F6a}.

\begin{figure}[htbp]
\begin{center}
\includegraphics[width=7cm]{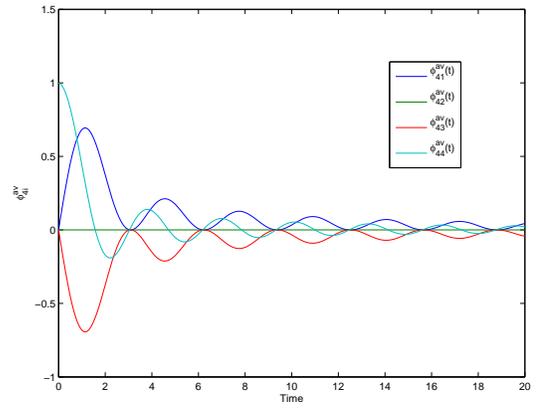}
\end{center}
\caption{Coefficient functions defining the time average of  $p_o(t)$.}
\label{F6b}
\end{figure}

To investigate the time average  property of the other quantum observer variable, we now plot the quantities
\begin{eqnarray*}
\phi_{41}^{ave}(T) &=& \frac{1}{T}\int_0^T\phi_{41}(t)dt\nonumber \\ 
\phi_{42}^{ave}(T) &=& \frac{1}{T}\int_0^T\phi_{42}(t)dt\nonumber \\ 
\phi_{43}^{ave}(T) &=& \frac{1}{T}\int_0^T\phi_{43}(t)dt\nonumber \\ 
\phi_{44}^{ave}(T) &=& \frac{1}{T}\int_0^T\phi_{44}(t)dt\nonumber \\ 
\end{eqnarray*}
in Figure \ref{F6b}. 

\begin{figure}[htbp]
\begin{center}
\psfrag{phi1i}{$\tilde\phi_{1i}$}
\psfrag{phi11t}{\tiny $\tilde\phi_{11}$}
\psfrag{phi12t}{\tiny $\tilde\phi_{12}$}
\psfrag{phi13t}{\tiny $\tilde\phi_{13}$}
\psfrag{phi14t}{\tiny $\tilde\phi_{14}$}
\includegraphics[width=7cm]{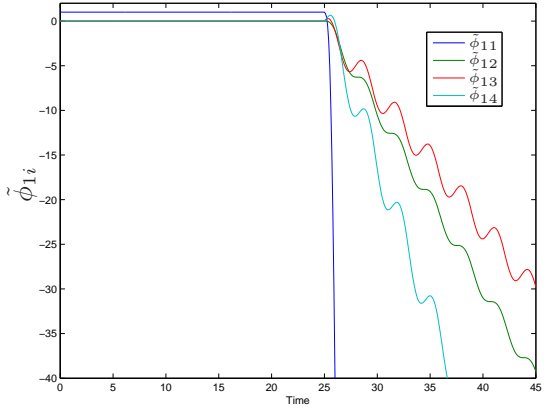}
\end{center}
\caption{Coefficient functions defining $q_p(t)$.}
\label{F7}
\end{figure}

We now illustrate the comments in Remark \ref{R2} by supposing that the above quantum observer is applied to the quantum plant for the time interval $t \in [0,20]$. Then, the quantum observer is disconnected from the quantum plant for the time interval $t \in [20,25]$. During this time internal, the quantum plant can be regarded to be connected to a null quantum observer so that $A_a = 0$ in this time interval. At time $t=25$, the quantum plant is then connected to a different quantum observer defined by the parameters $R_o = I$, $C_o = [0~1]$, $\beta = \left[\begin{array}{l}0\\1\end{array}\right]$  and $\alpha = \left[\begin{array}{l}0\\-1\end{array}\right]$. We write 
\begin{eqnarray*}
A_{a1}
&=& \left[\begin{array}{llll}      0  &   0  &  0  &   0 \\
     0 &    0  &   2  &   0 \\
    0 &    0  &   0  &   2 \\
     2 &    0  &   -2  &  0
\end{array}\right],\quad A_{a2} = 0, \nonumber \\
A_{a3}
&=& \left[\begin{array}{llll}      0  &   0  &  0  &   -2 \\
     0 &    0  &   0  &   0 \\
    0 &    -2  &   0  &   2 \\
     0 &    0  &   -2  &  0
\end{array}\right]
\end{eqnarray*}
so that the matrix $A_{a1}$ defines the dynamics of the augmented plant-observer system in the time interval $t \in [0,20]$, the matrix $A_{a2}$ defines the dynamics of the augmented plant-observer system in the time interval $t \in [20,25]$, and  the matrix $A_{a3}$ defines the dynamics of the augmented plant-observer system for $t \geq 25$. Then, we can write 
\[
x_a(t) = \tilde \Phi(t) x_a(0) 
\]
where 
\begin{eqnarray*}
 \tilde \Phi(t) &=& \left\{\begin{array}{l} e^{A_{a1} t} \mbox{ for } t \in [0,20],
 \\
  e^{A_{a2} (t-20)}e^{A_{a1} 20}= e^{A_{a1} 20}\mbox{ for } t \in [20,25],
\\
  e^{A_{a3} (t-25)}e^{A_{a1} 20}\mbox{ for } t \geq 25
\end{array} \right.\\
&=&\left[\begin{array}{llll}      \tilde\phi_{11}(t)  &   \tilde\phi_{12}(t)  &  \tilde\phi_{13}(t)  &  \tilde\phi_{14}(t) \\
     \tilde\phi_{21}(t)  &   \tilde\phi_{22}(t)  &  \tilde\phi_{23}(t)  &  \tilde\phi_{24}(t) \\
\tilde\phi_{31}(t)  &   \tilde\phi_{32}(t)  &  \tilde\phi_{33}(t)  &  \tilde\phi_{34}(t) \\
\tilde\phi_{41}(t)  &   \tilde\phi_{42}(t)  &  \tilde\phi_{43}(t)  &  \tilde\phi_{44}(t)
\end{array}\right].
 \end{eqnarray*}
Now in a similar fashion to Figure \ref{F3}, we plot the quantities $\tilde\phi_{11}(t)$,   $\tilde\phi_{12}(t)$, $\tilde\phi_{13}(t)$, and
$\tilde\phi_{14}(t)$ in Figure \ref{F7}.

\begin{figure}[htbp]
\begin{center}
\psfrag{phi2i}{$\tilde\phi_{2i}$}
\psfrag{phi21t}{\tiny $\tilde\phi_{21}$}
\psfrag{phi22t}{\tiny $\tilde\phi_{22}$}
\psfrag{phi23t}{\tiny $\tilde\phi_{23}$}
\psfrag{phi24t}{\tiny $\tilde\phi_{24}$}
\includegraphics[width=7cm]{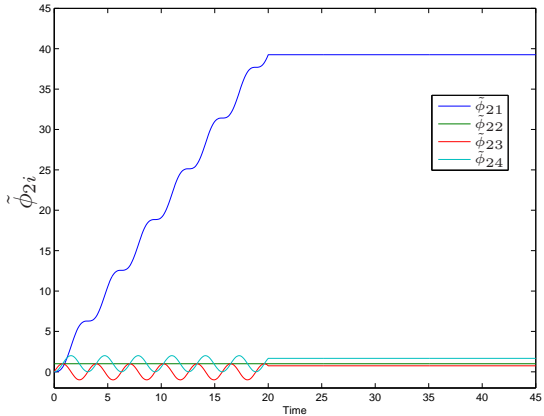}
\end{center}
\caption{Coefficient functions defining $p_p(t)$.}
\label{F8}
\end{figure}

Also, in a similar fashion to Figure  \ref{F4} we plot the quantities $\tilde\phi_{21}(t)$,   
$\tilde\phi_{22}(t)$, $\tilde\phi_{23}(t)$, and
$\tilde\phi_{24}(t)$ in Figure \ref{F8}.

\begin{figure}[htbp]
\begin{center}
\psfrag{phi3i}{$\tilde\phi_{3i}$}
\psfrag{phi31t}{\tiny $\tilde\phi_{31}$}
\psfrag{phi32t}{\tiny $\tilde\phi_{32}$}
\psfrag{phi33t}{\tiny $\tilde\phi_{33}$}
\psfrag{phi34t}{\tiny $\tilde\phi_{34}$}
\includegraphics[width=7cm]{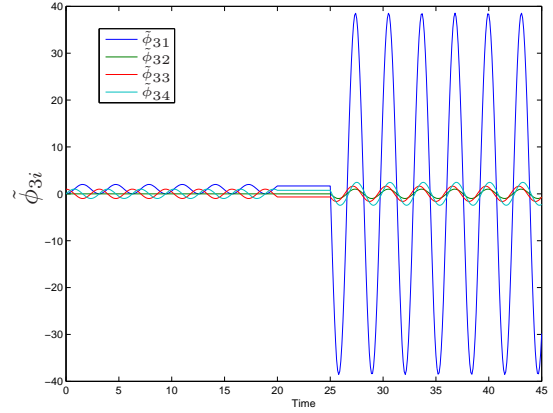}
\end{center}
\caption{Coefficient functions defining $q_o(t)$.}
\label{F9}
\end{figure}

Moreover, in a similar fashion to Figure  \ref{F5} we plot the quantities $\tilde\phi_{31}(t)$,   
$\tilde\phi_{32}(t)$, $\tilde\phi_{33}(t)$, and
$\tilde\phi_{34}(t)$ in Figure \ref{F9}.

In addition, in a similar fashion to Figure  \ref{F6}, we now plot the quantities
\begin{eqnarray*}
\tilde \phi_{31}^{ave}(T) &=& \frac{1}{T}\int_0^T\tilde \phi_{31}(t)dt\nonumber \\ 
\tilde \phi_{32}^{ave}(T) &=& \frac{1}{T}\int_0^T\tilde \phi_{32}(t)dt\nonumber \\ 
\tilde \phi_{33}^{ave}(T) &=& \frac{1}{T}\int_0^T\tilde \phi_{33}(t)dt\nonumber \\ 
\tilde \phi_{34}^{ave}(T) &=& \frac{1}{T}\int_0^T\tilde \phi_{34}(t)dt\nonumber \\ 
\end{eqnarray*}
in Figure \ref{F6}. 
\begin{figure}[htbp]
\begin{center}
\psfrag{phi3i}{$\tilde\phi_{3i}^{av}$}
\psfrag{phi31t}{\tiny $\tilde\phi_{31}$}
\psfrag{phi32t}{\tiny $\tilde\phi_{32}$}
\psfrag{phi33t}{\tiny $\tilde\phi_{33}$}
\psfrag{phi34t}{\tiny $\tilde\phi_{34}$}
\includegraphics[width=7cm]{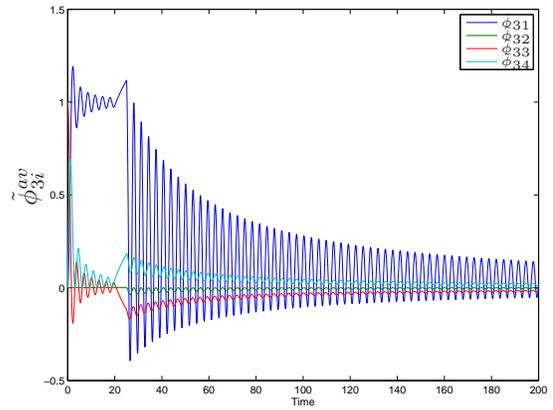}
\end{center}
\caption{Coefficient functions defining the time average of  $q_o(t)$.}
\label{F10}
\end{figure}

\begin{figure}[htbp]
\begin{center}
\psfrag{phi4i}{$\tilde\phi_{4i}$}
\psfrag{phi41t}{\tiny $\tilde\phi_{41}$}
\psfrag{phi42t}{\tiny $\tilde\phi_{42}$}
\psfrag{phi43t}{\tiny $\tilde\phi_{43}$}
\psfrag{phi44t}{\tiny $\tilde\phi_{44}$}
\includegraphics[width=7cm]{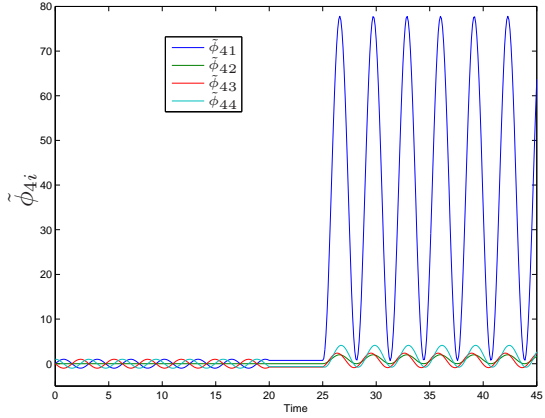}
\end{center}
\caption{Coefficient functions defining $p_o(t)$.}
\label{F11}
\end{figure}

Also, in a similar fashion to Figure  \ref{F6a}, we plot the quantities $\tilde\phi_{41}(t)$,   
$\tilde\phi_{42}(t)$, $\tilde\phi_{43}(t)$, and
$\tilde\phi_{44}(t)$ in Figure \ref{F11}.

\begin{figure}[htbp]
\begin{center}
\psfrag{phi4i}{$\tilde\phi_{4i}^{av}$}
\psfrag{phi41t}{\tiny $\tilde\phi_{41}$}
\psfrag{phi42t}{\tiny $\tilde\phi_{42}$}
\psfrag{phi43t}{\tiny $\tilde\phi_{43}$}
\psfrag{phi44t}{\tiny $\tilde\phi_{44}$}
\includegraphics[width=7cm]{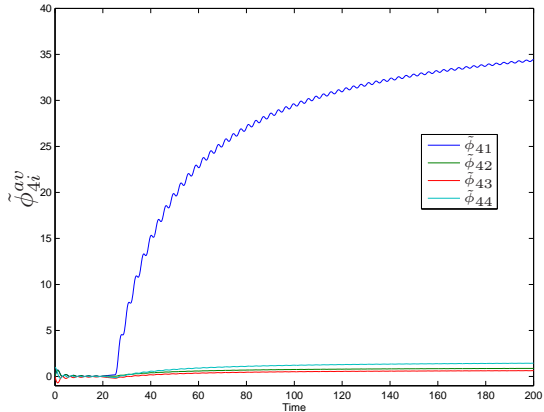}
\end{center}
\caption{Coefficient functions defining the time average of  $p_o(t)$.}
\label{F12}
\end{figure}

In addition, in a similar fashion to Figure  \ref{F6b}, we now plot the quantities
\begin{eqnarray*}
\tilde \phi_{41}^{ave}(T) &=& \frac{1}{T}\int_0^T\tilde \phi_{41}(t)dt\nonumber \\ 
\tilde \phi_{42}^{ave}(T) &=& \frac{1}{T}\int_0^T\tilde \phi_{42}(t)dt\nonumber \\ 
\tilde \phi_{43}^{ave}(T) &=& \frac{1}{T}\int_0^T\tilde \phi_{43}(t)dt\nonumber \\ 
\tilde \phi_{44}^{ave}(T) &=& \frac{1}{T}\int_0^T\tilde \phi_{44}(t)dt\nonumber \\ 
\end{eqnarray*}
in Figure \ref{F12}. 

\section{Conclusions}
In this paper we have introduced a notion of a direct coupling observer for closed quantum linear systems and given a result which shows how such an observer can be constructed. The main result shows the time average convergence properties of the direct coupling observer. We have also presented an illustrative example along with simulations to investigate the behavior of a direct coupling observer when applied to a simple one mode quantum linear system. Future research in this area might involve extending the class of quantum linear systems for which a direct coupling observer can be designed and also considering the problem of constructing an observer which is optimal in some sense. Also, future research could investigate the role of direct coupling observers in the design of direct coupling coherent quantum feedback control systems. 

\section*{Acknowledgement}
The author would like to thank Dr. Igor Vladimirov for his useful comments.


\end{document}